\documentclass[10pt,conference]{IEEEtran}
\IEEEoverridecommandlockouts
\usepackage{cite}
\usepackage{amsmath,amssymb,amsfonts}
\usepackage[numbers]{natbib}
\usepackage{graphicx}
\usepackage{textcomp}
\usepackage{xcolor}
\usepackage{balance}

\usepackage[hidelinks]{hyperref}
\usepackage[paperVersion]{dpupaper}
\usepackage{subcaption}

\def\BibTeX{{\rm B\kern-.05em{\sc i\kern-.025em b}\kern-.08em
    T\kern-.1667em\lower.7ex\hbox{E}\kern-.125emX}}
\begin{document}

\title{Fast and Memory-Efficient Neural Code Completion
\thanks{$^\diamond$Work done as AI Residents in Microsoft Research, Cambridge, UK. Equal contribution.}}

\author{\IEEEauthorblockN{Alexey Svyatkovskiy\IEEEauthorrefmark{1},
 Sebastian Lee\IEEEauthorrefmark{2}$^\diamond$,
 Anna Hadjitofi\IEEEauthorrefmark{3}$^\diamond$,\\
 Maik Riechert\IEEEauthorrefmark{1},
 Juliana Vicente Franco\IEEEauthorrefmark{1},
 Miltiadis~Allamanis\IEEEauthorrefmark{1}}
\IEEEauthorblockA{\IEEEauthorrefmark{1}\textit{Microsoft}\\
\{alsvyatk,marieche,jufranc,miallama\}@microsoft.com}
\IEEEauthorblockA{\IEEEauthorrefmark{2}\textit{University of Oxford}\\
sebalexlee@gmail.com}
\IEEEauthorblockA{\IEEEauthorrefmark{3}\textit{University of Edinburgh}\\
a.hadjitofi@ed.ac.uk}
}

\maketitle

\begin{abstract}
  Code completion is one of the most widely used features of modern integrated development environments (IDEs).
  While deep learning has made significant progress in the
  statistical prediction of source code, state-of-the-art
  neural network models consume hundreds of megabytes
  of memory, bloating the development environment.
  We address this in two steps: first we present a modular neural framework for code completion.
  This allows us to explore the design space and evaluate different techniques.
  Second, within this framework we design a novel reranking neural completion
  model that combines static analysis with granular token encodings.
  The best neural reranking model consumes just 6 MB of RAM, --- 19x less
  than previous models ---
  computes a single completion in 8 ms, and achieves 90\% accuracy in its top
  five suggestions.
\end{abstract}

\begin{IEEEkeywords}
code completion, deep learning, API completion
\end{IEEEkeywords}

%

\section{Introduction}
\label{sect:introduction}
Deep learning has a substantial impact on software engineering
methods across a variety of tasks~\citep{allamanis2018survey}.
One early application of machine learning of source code
has been code completion~\citep{bruch2009learning,hindle2012naturalness,nguyen2013statistical}, \ie
the suggestion of code that a developer is about to
type. Code completion is the most frequently used feature in IDEs~\citep{amann2016study,murphy2006java}.
Early machine learning methods
used feature-based models to predict the next function to be invoked~\citep{bruch2009learning,proksch2015intelligent}. 
Other models, such as $n$-gram language models~\citep{hindle2012naturalness},
do not require extraction of features, but instead use the code's tokens
and --- to some extent --- can generalize to new code and APIs~\citep{hellendoorn2017deep}.
However, $n$-gram models have a prohibitive footprint consuming gigabytes of RAM~\citep{karampatsis2019maybe}.

Recently, neural machine learning methods
have been found to be effective for code completion, achieving
state-of-the-art accuracy~\citep{karampatsis2019maybe,svyatkovskiy2019pythia,aye2020sequence}.
Nevertheless, these models commonly have a large memory and computational footprint --- consuming 
many hundreds of megabytes in RAM, which is prohibitive for any single component
of a modern IDE with hundreds of features.
Furthermore, despite the promising results from neural code completion models,
they fail to capture rich information from static
analyses (\eg type information) that is directly useful to
code completion.

In this work, we tackle these limitations by reformulating neural
code completion from \emph{generation} to \emph{ranking}.
We achieve this by taking advantage of the candidate completion suggestions generated by
pre-existing static analyses. This improves the predictive
accuracy compared to baseline models and enables us to use fine-level
encodings of code tokens, which reduce or completely remove the
need for maintaining a memory-intensive vocabulary and embedding matrix, 
while achieving good accuracy trade-offs.
We create efficient neural code completion models that consume just
6 MB of RAM --- 19$\times$ less than the baselines --- and execute in a few milliseconds while
still achieving 90\% accuracy in their
top five suggestions for API completion. Such systems can support a wide
variety of realistic developer environments, including those that are significantly resource-constrained,
and avoid introducing bloat to editors and IDEs.
This is important as we seek to provide
inclusive tools to developers --- including those without access to high-end
machines or Internet connections.

In brief,
(a) we present a modular neural code completion framework that allows us to
explore a wide range of neural components with varying trade-offs in terms of
memory, speed, and accuracy (\autoref{sect:approach});
(b) within this framework, we present a novel reranking neural model that uses static analysis
and granular neural token encodings combining the best of static analysis-based code
completion and neural-based code completion (\autoref{sect:approach});
(c) we implement our framework for API completion and
present an extensive and principled evaluation over the design space (\autoref{sect:evaluation}),
testing 64 model configurations across multiple hyperparameters spending more than one GPU-year
of computational effort. Our evaluation shows that our novel neural reranking models
have a memory footprint as little as 6~MB, achieving 90\% completion accuracy 
on the top five suggestions within 8~ms.
The studied code completion models are now used in Microsoft's Visual Studio 
IntelliCode.

\section{Approach}
\label{sect:motivating}
\begin{figure}
   \centering
   \includegraphics[width=0.95\columnwidth]{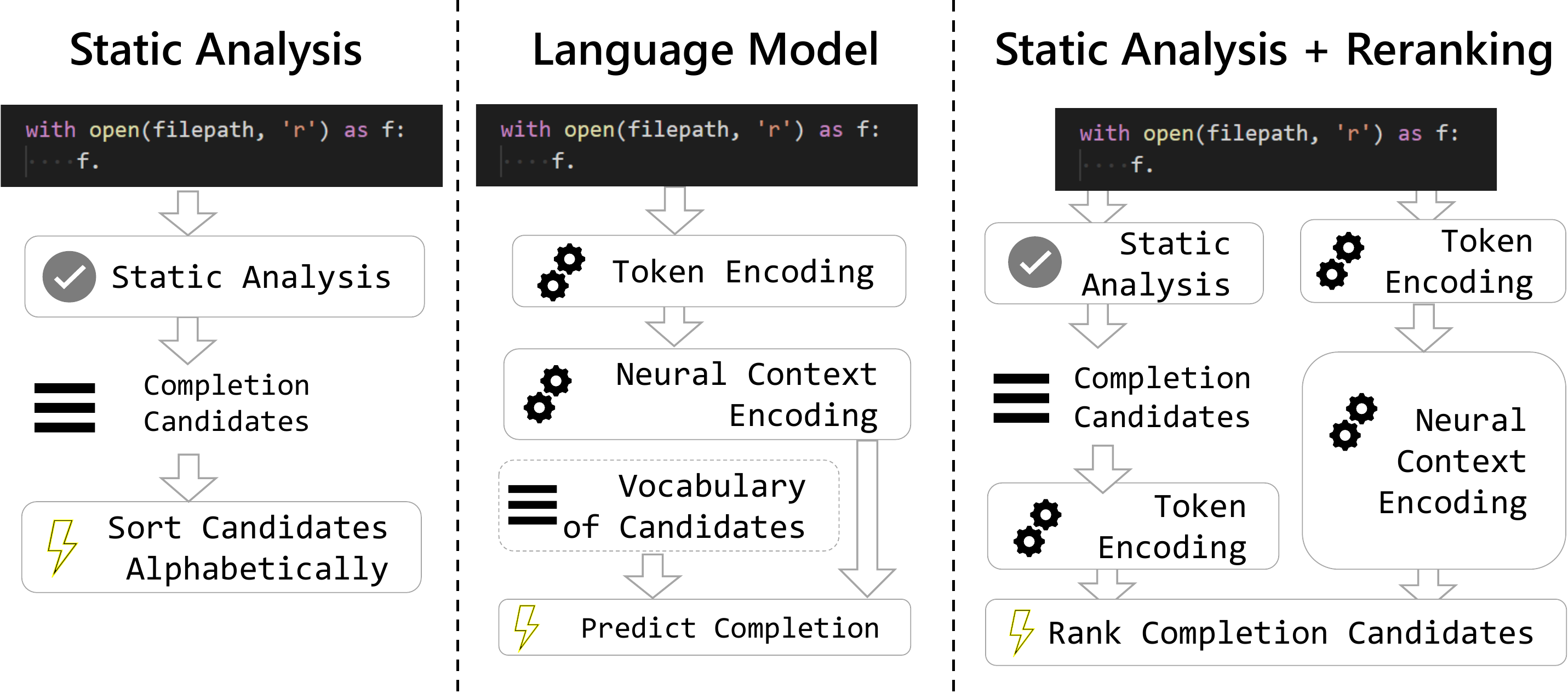}
   \caption{Static Analysis-based code completion (left), 
      Language model-based code completion (center),
      Neural rerank models with static analysis (this work; right).}\label{fig:design increments}
\end{figure}

\begin{figure}
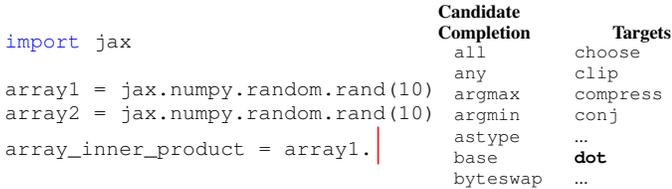

   \begin{minipage}{0.6\columnwidth}
      \begin{lstlisting}[language=Python]
import jax

array1 = jax.numpy.random.rand(10)
array2 = jax.numpy.random.rand(10)
array_inner_product = array1.(*{\color{red}\Large$\mid$}*)
      \end{lstlisting}
   \end{minipage}
   ~~
   \begin{minipage}{0.35\columnwidth} \scriptsize
      \textbf{Candidate Completion Targets}
      \begin{tabular}{ll}
         \code{all} & \code{choose}\\
         \code{any} & \code{clip}\\
         \code{argmax} & \code{compress} \\
         \code{argmin} & \code{conj} \\
         \code{astype} & ... \\
         \code{base} & \textbf{\texttt{dot}} \\
         \code{byteswap} & ... \\
      \end{tabular}
   \end{minipage}

   \caption{Motivating Example. A developer using \href{https://github.com/google/jax}{\code{jax}}
            is declaring \code{array\_inner\_product} using the \code{jax} API (left; cursor in red).
            IDE completion tools ask a completion model to return a ranked list of
            \emph{candidate completion targets} at the \emph{completion location} which are
            then shown to the user (right).} \label{fig:motivating_jax}
\end{figure}

We use the synthetic snippet in \autoref{fig:motivating_jax} as a running example.
A developer is currently declaring the \code{array\_inner\_product} variable. We call the
current location of the cursor (shown in red) the \emph{completion location}, which is where
a code completion tool has the opportunity to serve \emph{completion suggestions}.
Commonly, an IDE will perform a static analysis (\autoref{fig:design increments}; left) to determine the type of \code{array1}
and return the list of \emph{candidate completions} that are valid at the
given location and the given \emph{completion context} (declared variables, imported packages, \etc).
Traditionally, IDEs did not employ learned components and instead yielded an
alphabetically sorted list of type-correct suggestions. In this case, the list of suggestions
is quite long (\autoref{fig:motivating_jax}; right). One approach, used by
\citet{proksch2015intelligent,bruch2009learning}, is to extract
hard-coded features from the completion context and learn which candidate
completion targets are relevant in the given context. This approach, however,
misses the opportunity to learn richer features directly from data --- an
approach that has recently been made possible thanks to deep learning. For example,
the name of the variable \code{array\_inner\_product} indicates that the developer is
about to invoke the \code{dot} method to compute the inner product.
Manually anticipating and extracting relevant features
that cover as many cases as possible is difficult. Neural methods aim to address this.

Furthermore, such approaches learn about individual APIs and
cannot generalize to unseen ones.
This requires sufficient example
usages of an API to \emph{learn} about
those features. However, this is not always possible. In our
example, \code{jax} is a relatively new machine learning library that
is under active development and relatively few public
codebases use it. Most existing code completion methods
would not generalize to \code{jax} or other previously unseen completion targets
since they require sufficient training data of the API usage in real code.

Neural models alleviate this as they can
generalize better to previously unseen code by automatically learning
the aspects that are relevant for a given completion context.
Such features include the structure of the code, but also the
names of the variables and functions appearing in the context.
For \code{jax} (\autoref{fig:motivating_jax}), which is built to match \code{numpy} in some aspects,
neural models can recognize similarities in the
numeric manipulations in \autoref{fig:motivating_jax}.
An important source of information is contained in the names
within the completion context (\eg the subtokens in \code{array\_inner\_product} of \autoref{fig:motivating_jax}).
Early models~\citep{hindle2012naturalness,allamanis2013mining,tu2014localness,hellendoorn2017deep,bielik2016phog,maddison2014structured}
did \emph{not} take into account the structure within names, which
nevertheless contain valuable information. Only recently,
\citet{karampatsis2019maybe} introduced
such techniques. To advance this idea,
we devise a framework (\autoref{sect:approach}) that allows
us to test multiple techniques for learning about the
internal structure of names and show how these techniques
provide different trade-offs in terms of completion accuracy,
memory usage, and computational cost.

Furthermore, all existing neural models treat code completion as a
language modeling problem, \ie models are tasked with
\emph{generating} the target completion from their internal
knowledge (\autoref{fig:design increments}; center). Considering the diverse nature of the completion
targets, this is an unnecessarily hard task, since language models need to be aware of all
valid completions (\eg candidate completion targets
in \autoref{fig:motivating_jax}), or be able to reconstruct them from scratch.
Within our framework,
we present a novel model that treats the neural code completion
as the problem of \emph{(re)ranking} the candidate completion targets
returned from a static analysis (\autoref{fig:design increments}; right).
This advances the state-of-the-art
allowing us to build and deploy memory-efficient code completion tools
without sacrificing accuracy.

\section{A Neural Code Completion Framework}
\label{sect:approach}
\newcommand{\candidateProvider}{\ensuremath{\mathcal{P}}\xspace}
  \newcommand{\vocabProvider}{\textsc{Vocab}\xspace}
  \newcommand{\staticAnalysisProvider}{\textsc{StAn}\xspace}

\newcommand{\token}{\ensuremath{t}\xspace}
\newcommand{\tokenEncoder}{\ensuremath{\mathcal{E}}\xspace}
  \newcommand{\tokenUnitEncoder}{\textsc{Token}\xspace}
  \newcommand{\subtokenEncoder}{\textsc{Subtoken}\xspace}
  \newcommand{\bpeEncoder}{\textsc{bpe}\xspace}
  \newcommand{\charEncoder}{\textsc{Char}\xspace}

\newcommand{\contextEncoder}{\ensuremath{\mathcal{C}}\xspace}
  \newcommand{\rnnCxEncoder}{\textsc{Gru}\xspace}
  \newcommand{\birnnCxEncoder}{\textsc{biGru}\xspace}
  \newcommand{\cnnCxEncoder}{\textsc{Cnn}\xspace}
  \newcommand{\transformerCxEncoder}{\textsc{Transformer}\xspace}

\newcommand{\completionRanker}{\ensuremath{\mathcal{R}}\xspace}
\newcommand{\dotRanker}{\textsc{Dot}\xspace}

\newcommand{\modelName}[3]{\ensuremath{\left\langle #1,#2,#3\right\rangle}}

\begin{figure*}[t]
  \centering
  \includegraphics[width=0.99\textwidth]{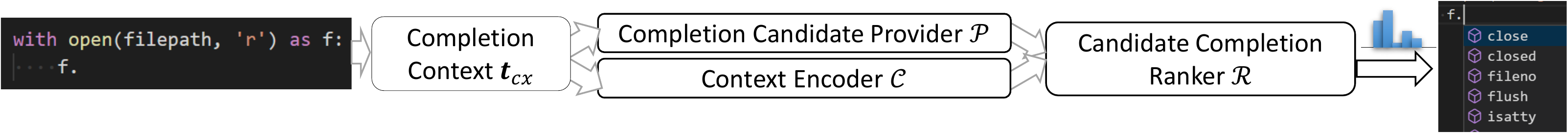}
    \caption{Our framework for machine learning-based code
             completion systems. From the developer environment, the (partial) code context
             $\vect{t}_{\text{cx}}$ is extracted and passed into the context encoder, \contextEncoder, that
             ``summarizes'' the completion context. Simultaneously, a candidate provider \candidateProvider
             receives the code context and computes a set of candidate completions for
             the current location. Finally, a completion ranker \completionRanker
             ranks the candidates given the summarized context and presents them to the user.
             In this work, we instantiate each component with different types of neural networks (\autoref{tbl:neural components}),
             showing how different choices affect suggestion accuracy,
             model size and computational cost.
             All neural models are trained end-to-end.
             } \label{fig:architecture}
\end{figure*}

First, we present a general neural framework
(\autoref{fig:architecture}).
Designing neural networks for a specific task is an engineering
effort that commonly involves combining different components (also
referred to as ``modules'') such that the neural network achieves
sufficient accuracy while complying to other non-functional requirements,
such as computational speed and memory consumption.
Here, we design a framework that allows us to perform a principled exploration of a series of design decisions
that can help us pick a good trade-off among the desired properties of
a practical completion system. Our framework distinguishes the
following components (\autoref{fig:architecture}):
\begin{description}
  \item[Token Encoder] A neural network \tokenEncoder encoding
    a code token $t$ into a distributed
    vector representation (embedding) $\vect{r}_t$.
  \item[Context Encoder] A neural network $\contextEncoder$ that encodes
    (\ie summarizes) the completion context $\vect{t}_{\text{cx}}$, into a distributed vector
    representation (embedding) $\vect{c}_{\text{cx}}$.
  \item[Candidate Provider \candidateProvider] A component that accepts the completion
    context $\vect{t}_{\text{cx}}$ and yields a set of $M$ candidate completion targets $s_i,$
    \ie $\candidateProvider(\vect{t}_{\text{cx}}) = \{s_i\} = \{s_0, ..., s_M\}$.
  \item[Completion Ranker] A neural component $\completionRanker$
    that accepts the context encoding $\vect{c}_{\text{cx}}$, along with a set of candidate
    completion targets $\{s_i\},$ and ranks them.
\end{description}
The components, their inputs/outputs, and concrete implementations are shown in \autoref{tbl:neural components}.
We denote a concrete configuration with a tuple of the form \modelName{\tokenEncoder}{\contextEncoder}{\candidateProvider}.
For example, \modelName{\subtokenEncoder}{\rnnCxEncoder}{\staticAnalysisProvider} is an instantiation of the
framework with a \subtokenEncoder token encoder \tokenEncoder, a \rnnCxEncoder context encoder \contextEncoder
and a \staticAnalysisProvider completion provider \candidateProvider.
Our architecture is general and subsumes most
neural language models~\citep{hellendoorn2017deep,karampatsis2019maybe,svyatkovskiy2019pythia,aye2020learning}.
By selecting a concrete component implementation, we get
a range of neural code completion models.
To restrict the explored space, we fix two aspects.
Following current state-of-the-art~\citep{karampatsis2019maybe}, we
treat the completion context as a list of the $N$ tokens before the completion location,
\ie $\vect{t}_{\text{cx}} = [ \token_0, ..., \token_{N-1}]$.
For example, the last four elements of the completion context of \autoref{fig:motivating_jax} are
[\code{array\_inner\_product}, \code{=}, \code{array1}, \code{.}].
Although other representations of code have been explored~\citep{allamanis2018survey}
and could be used,
token-based models have been shown to achieve good performance
in code completion~\citep{karampatsis2019maybe}, with a small computational
cost for extracting information from the code context.
For example, the graph representations of \citet{allamanis2018learning}
contains significantly more information, but the computational cost of processing these structures is
prohibitive for real-time systems. We thus pass to the context encoder
$\contextEncoder$ the context $\vect{t}_{\text{cx}}$ and the token encoder \tokenEncoder.
The second design decision is to tie the token encoder within
\contextEncoder and \completionRanker. This is a common and theoretically
principled architectural choice that yields improved results,
while reducing memory requirements~\citet{inan2016tying}.
In the rest of this section, we discuss concrete implementations
for each component.

\begin{table*}[tb]
  \caption{Components of the framework in \autoref{fig:architecture}. By combining these components,
  we retrieve a concrete neural code completion model system. We denote a specific architecture
  using a tuple, \eg \modelName{\subtokenEncoder}{\rnnCxEncoder}{\staticAnalysisProvider}
  is a model with a static analysis-based candidate provider \candidateProvider, a subtoken-based token encoder
  \tokenEncoder, and an RNN-based context encoder \contextEncoder.}\label{tbl:neural components}
  \begin{tabular}{lllp{8cm}}
  \toprule
    Component           & Signature           & Returns           & Implementations \\ \midrule
    Token Encoder       & $\tokenEncoder(t)$  & Token Embedding $\vect{r}_t \in \mathbb{R}^D$    & \tokenUnitEncoder, \subtokenEncoder, \bpeEncoder, \charEncoder \\
    Context Encoder     & $\contextEncoder(\vect{t}_{\text{cx}}, \tokenEncoder)$  & Context Embedding $\vect{c}_{\text{cx}} \in \mathbb{R}^H$ & \rnnCxEncoder, \birnnCxEncoder, \transformerCxEncoder, \cnnCxEncoder, and annotated ($\diamond$) variants\\
    Candidate Provider    & $\candidateProvider(\vect{t}_{\text{cx}})$  & Candidate Completions Set $\{s_i\}$ & \vocabProvider, \staticAnalysisProvider \\
    Completion Ranker   & $\completionRanker(\tokenEncoder, \{s_i\}, \vect{c}_{\text{cx}})$ & Ranked suggestions by $P(s_i|\vect{c}_{\text{cx}})$ &  \dotRanker \\
  \bottomrule
  \end{tabular}
\end{table*}

\subsection{Token Encoders \tokenEncoder}\label{subsection:token_encoder}
There is a wide literature about encoding
tokens in source code and text. A key characteristic of source code tokens
is that they tend to be extremely sparse,
combining different words (commonly called ``subtokens'') to create a single code token~\citep{allamanis2013mining}.
For example \code{array\_inner\_product} is a very rare name, but is
made of three more common subtokens. We consider four commonly
employed token encoders. All of the presented token encoders have been used in
some form in previous works or in natural language processing (NLP)
and offer alternative trade-offs in terms of their ability to represent tokens,
memory requirements, and computational cost.

\boldpara{Token-Unit Encoder (\tokenUnitEncoder)}
  The simplest and most commonly used token encoder.
  \tokenUnitEncoder learns an embedding of dimension $D$ for each token in a fixed vocabulary $V_t$. This
  requires learning and storing an embedding matrix with $|V_t| \times D$ parameters.
  \tokenUnitEncoder then performs a lookup, \ie
  \begin{align} \label{eq:tokenEncoder}
     \tokenEncoder_{\tokenUnitEncoder}(t) = \textsc{EmbeddingLookup}\left(t, V_t\right),
  \end{align}
  where $\textsc{EmbeddingLookup}(t, V_t)$ returns the $D$-dimensional row
  of the embedding matrix that corresponds to $t$.
  If the lookup fails, then the learned embedding of a special unknown identifier
  (``\textsc{Unk}'') is returned. The vocabulary $V_t$ is selected from the training data
  and contains the most frequent tokens and the \textsc{Unk} symbol. The size of the
  vocabulary $\vert V_t\vert$ is a hyperparameter that needs to be tuned: smaller vocabularies reduce
  memory usage at the cost of failing to represent many tokens and thus yielding
  less accurate suggestions.
Commonly, \tokenUnitEncoder has a large number of parameters:
  for practical vocabulary sizes and sufficiently expressive embedding dimension $D$,
  the number of parameters is in the order
  of $10^7$, which is orders of magnitude more than the number of parameters typically
  required for context encoders and amounts to many megabytes which need
  to be stored in the RAM.

\boldpara{Subtoken Encoder (\subtokenEncoder)}
  Source code identifiers are often made up of smaller parts.
  For example, \code{array\_inner\_product} is made up of three subtokens
  (\code{array}, \code{inner}, \code{product}). \subtokenEncoder learns
  to \emph{compose} the meaning of an identifier from its subtokens into
  a single encoding. This allows to better capture the sparse nature of
  identifiers while simultaneously reducing the memory requirements compared
  to \tokenEncoder.
  \subtokenEncoder sub-tokenizes identifiers and embeds them
  using an embedding matrix with size $|V_s| \times D$, where $V_s$ is the
  subtoken ``vocabulary''. Since subtokens are less sparse
  than tokens, $|V_s| \ll |V_t|,$ and thus \subtokenEncoder
  affords a smaller embedding matrix.
  Obtaining a representation of a token $t$ requires composing the
  representation from the subtoken embeddings of the token, \ie
  \newcommand{\aggOp}{\ensuremath{\bigoplus}}
  \begin{align}\label{eq:subtokenEncoder}
      \tokenEncoder_{\subtokenEncoder}(t) = \aggOp_{t_s \in \textsc{Split}(t)} \textsc{EmbeddingLookup}\left(t_s, V_s\right),
  \end{align}
  where $\textsc{EmbeddingLookup}(\cdot)$ is defined analogously to the word-level case, 
  \textsc{Split($\cdot$)} is a deterministic function that subtokenizes a string on
  camelCase and pascal\_case yielding a list of lower-cased subtokens.
  $\oplus$ is a permutation-invariant aggregation operator
  that ``summarizes'' the subtoken embeddings. We tested three
  options for $\oplus$: element-wise summation, average, and maximum.
  In early experiments, all operators achieved comparable performance; for conciseness we
  only report results for the elementwise maximum operator.

  \boldpara{BPE Encoder (\bpeEncoder)}
    Byte-pair encoding (BPE) is a commonly used NLP method
    for dealing with rare words~\citep{sennrich2016neural} with
    origins in data compression. Specifically, BPE uses a preprocessing step to ``learn''
    subtokens by combining commonly occurring consecutive characters. Then each token
    is represented as a sequence of those subtokens. Note that the way that a token is
    split depends on the training data and is ``learned'' during preprocessing.
    For example, \bpeEncoder may split \code{array\_inner\_product} into
    \code{array}, \code{\_}, \code{in}, \code{ner}, \code{\_}, \code{prod}, \code{uct}.
    Our \bpeEncoder encoder is identical to
    \subtokenEncoder but replaces \textsc{Split} in \autoref{eq:subtokenEncoder}
    with the BPE-based splitting.

  \boldpara{Character-Based Encoder (\charEncoder)}
  Finally, we consider a character-level encoder. \charEncoder
  composes a representation of a token from its individual characters.
  The main benefits of \charEncoder is that it commonly has a significantly smaller
  number of parameters compared to other encoders and that it can represent
  arbitrary tokens as long as they are made up from known characters. Token 
  representations are then \textit{computed}. Thus, \charEncoder stores \emph{only}
  the network parameters and has no vocabulary. The trade-off
  is that \charEncoder commonly have lesser representational capacity
  and is more computationally expensive to the other encoders.
  For a token $t$,
  \begin{align}
  \tokenEncoder_{\charEncoder}(t)=\textsc{1dCnn}(\textsc{GetChars}(t)),
  \end{align}
  where \textsc{1dCnn} is a 1D convolutional neural network (CNN) and
  \textsc{GetChars} splits $t$ into a list of characters. The \textsc{1dCnn} is
  similar to previous NLP work~\citep{zhang2015character,kim2016character}.
  We create an alphabet of the commonly used characters.
  Each token is then represented as a matrix of
  one-hot columns with the element corresponding to the relevant index in
  the alphabet set to 1.

\subsection{Context Encoders \contextEncoder}\label{context_encoder}
Context encoders (\autoref{fig:architecture}, \autoref{tbl:neural components}) accept the completion context
and encode all information that
is relevant at the current completion location into a single vector
representation $\vect{c}_{\text{cx}}$. Again, there is a large set of
design options. For computational efficiency we only
consider context encoders of the form
\begin{align} \label{eq:cx encoding}
  \vect{c}_{\text{cx}} = \contextEncoder\left(\vect{t}_{\text{cx}}, \tokenEncoder\right) = \contextEncoder\left(\tokenEncoder(\token_0), ..., \tokenEncoder(\token_{N-1})\right),
\end{align}
\ie context encoders that accept as input the $N$ context
tokens before the completion location and a token encoder \tokenEncoder.
The output $\vect{c}_{\text{cx}}$ is an $H$-dimensional
vector, where $H$ is a hyperparameter. One benefit of encoders of this form at
test-time is that the token encodings can be cached.

\boldpara{RNN-based Context Encoders} (\rnnCxEncoder)
Recurrent neural networks (RNN) are a common neural module that
summarize variable-length sequences. RNN encoders take
the form
\begin{align}
  \vect{h}_\nu = \textsc{RnnCell}\left(\vect{h}_{\nu-1}, \tokenEncoder(t_{\nu-1})\right),
\end{align}
where $\vect{h}_{\nu}$ is the vector state at position $\nu$,
$\tokenEncoder(t_{\nu})$ is the encoding of the input at position $\nu$ and
\textsc{RnnCell} is a learnable function. We test two commonly
used \textsc{RnnCell}s, LSTMs~\citep{hochreiter1997long} and GRUs~\citep{bahdanau2014neural}.
Given their similar performance, we only report results for GRUs.
We also test a bi-directional GRU, denoted as \birnnCxEncoder.

\boldpara{CNN Context Encoders} (\cnnCxEncoder)
Similar to the \charEncoder token encoder, 1D CNNs can be used
to encode the context. A
multitude of CNN configurations can be used.
All of them accept as an input an $N \times D$ matrix, and
after a few layers of convolution and pooling, a final pooling layer is used to
compute the context representation $\vect{c}_{\text{cx}}$.
This architecture has resemblance to the natural language classification
of \citet{kim2014convolutional}.

\boldpara{Transformer Context Encoders} (\transformerCxEncoder)
An alternative to RNN-based and CNN-based sequence models are
transformers~\citep{vaswani2017attention} which have recently shown exceptional
performance in NLP~\citep{devlin2018bert,radford2019language}.
Although transformers can be
parallelized efficiently on GPUs, they have quadratic runtime memory requirements with respect to
the sequence length. Here, we use the standard transformer encoder
architecture.

\boldpara{Completion Location-Annotated Encoders}
We can provide additional information to any of the above
context encoders \contextEncoder, \eg information derived from analyzing the code.
Here, we test adding lightweight information that is useful
for API completion. Specifically, we annotate all occurrences of the variable or namespace
on which an API completion is performed.
For example, in \autoref{fig:motivating_jax} we
indicate to the context encoders all the tokens that refer
to the object \code{array1}. This may allow a context encoder
to better recognize API patterns, \eg if \code{foo.open()} was
previously invoked then invoking \code{read()} on \code{foo} is likely.
Other annotations are also possible,
but we do \emph{not} test them in this work.

To capture this long-range context, we wrap the token
encoder of \autoref{eq:cx encoding} to append
a 0/1 component to each token encoding $\vect{r}_t$. This bit is set to
one for the tokens that are bound to the variable or
namespace whose API is about to be completed.
This provides additional information to the context encoders
at a minimal cost. We denote such encoders by appending a diamond ($\diamond$)
to their name, \eg $\rnnCxEncoder^\diamond$.

\subsection{Candidate Providers \candidateProvider}
We consider two types of candidate providers. \vocabProvider providers --- commonly used in
language model-based completion engines --- have a fixed list (vocabulary)
of candidate completions. The vocabulary is compiled from the
training data by taking the top most frequent target completions up to some size
$\mathbb{V}_{max}$, which is a model hyperparameter. Commonly,
the vocabulary is identical to $V_t$ defined by the \tokenUnitEncoder encoder.
The second provider is a static analysis-based provider (\staticAnalysisProvider).
Such providers are common in both typed and untyped languages
where a static analysis tool can determine a set of plausible
completions. For example, IntelliSense~\citep{intellisense},
PyCharm~\citep{pycharm}, and IntelliJ~\citep{intellij}
return only type-correct completions for a suggestion location.

These two providers offer different trade-offs:
\staticAnalysisProvider rely on static analyses and inherit their characteristics.
Commonly sound analyzers yield more precise and informative candidate completion targets
compared to \vocabProvider providers. Such candidate providers are preferred
by IDEs~\citep{svyatkovskiy2019pythia} since they often do \emph{not} make suggestions that are invalid
at the completion location which may confuse developers. Nevertheless, 
design decisions and heuristics when creating the underlying static analysis may hurt 
recall. Furthermore, \vocabProvider providers
can function in partial or incomplete contexts where static analyses cannot yield informative results.
Although using a predefined vocabulary of completions simplifies the machine learning
model, it does not allow for the model to provide candidates
beyond those seen in the training data; a major limitation for
suggesting and generalizing to rare, evolved or previously unseen APIs, such
as the \code{jax} API in \autoref{fig:motivating_jax}.
Although  a \vocabProvider provider can use a static analysis-based post-processing
step to filter some false positives it cannot generalize beyond its fixed vocabulary.

\boldpara{Other Candidate Providers}
Beyond \vocabProvider and \staticAnalysisProvider, other candidate providers can be used.
One commonly used case is ``structural prediction'' providers that learn to predict
completion targets by composing them from individual
subunits, such as subtokens~\citep{allamanis2015suggesting}, BPE~\citep{karampatsis2019maybe}
or characters, allowing for an unbounded set of possible candidates.
This is currently used in state-of-the-art work of \citet{karampatsis2019maybe}
and \citet{aye2020sequence}.
We do not test such providers since \staticAnalysisProvider-based providers
have access to strictly more information and a smaller hypothesis space to predict from.
Thus models with similar architecture but  with \staticAnalysisProvider 
providers will perform better in comparison with structural prediction
providers~\citep{pratt1995introduction}.
Furthermore, the generation of
candidate completions requires multiple steps (\eg beam search requiring
one step per subtoken/character) imposing additional
computational burden and
makes the generation of invalid completions more probable.

\subsection{Completion Ranker \completionRanker}
We only test \dotRanker,  a commonly used ranker~\citep{mitra2018introduction}.
\dotRanker ranks a
set of candidate completion targets, $\{s_i\}$, according to the probability
distribution
\begin{align}
  P(s_k|\vect{c}_{\text{cx}}, \{s_i\}, \tokenEncoder) =
      \frac{\exp\left((W\vect{c}_{\text{cx}})^\top \tokenEncoder(s_k) + b_{s_k} \right)}{\sum_{s_j \in \{s_i\}}\exp\left((W\vect{c}_{\text{cx}})^\top \tokenEncoder(s_j) + b_{s_j}\right)}, \label{eqn:prob}
\end{align}
\ie a softmax over the dot product of the token encodings of candidate suggestions
with a linearly transformed context encoding $\vect{c}_{\text{cx}}$. Here, $W$ is a
linear layer of size $H \times D$, which is learned along with the rest of the model parameters,
and linearly maps the $H$-dimensional vector into a $D$-dimensional one.
The vector $\vect{b}$ is a learned bias signifying the ``popularity'' of a given
method independent from its context. Since we want our model to generalize
to APIs that were previously unseen, we set  $\vect{b}=\vect{0}$ for all non-\vocabProvider-based models.

When \autoref{eqn:prob} is used in conjunction
with a \vocabProvider-based candidate provider and a \tokenUnitEncoder encoder,
it reduces to a standard token-level language model.
However, when the candidate provider yields a variable-sized, context-dependent set of candidates,
(as \staticAnalysisProvider models do)
\autoref{eqn:prob} becomes a ranking model.
As we will explore in the next sections
ranking a smaller set of valid candidates is a simpler
problem than predicting a target from a large vocabulary,
leading to improved performance and smaller memory footprint.

\subsection{Composing Components: Model Zoo}
\autoref{tbl:neural components} summarizes the components and implementations discussed.
To create a code completion system, one
needs to pick an implementation for each component, and instantiate
a neural network. This leads to 64
combinations with varying accuracy, memory requirements, and computational costs.
Additionally, each component
has its own hyperparameters, yielding a large search space.
For a given configuration and hyperparameters each neural network
is trained by jointly optimizing its parameters (embedding matrices, layer weights, \etc) end-to-end
with the objective to minimize the cross entropy loss of \autoref{eqn:prob}.
Note that candidate providers do \emph{not} have any learnable parameters.
In our training, we employ early stopping and set the context size to $N=80$ tokens,
which represents a reasonable trade-off between the amount of context and computation/memory
needed.

\section{Evaluation}
\label{sect:evaluation}
In the previous section, we presented a general framework for neural models
of code completion. To evaluate the multitude of configurations,
we focus on a particular instance of code completion, API completion. This
is the suggestion of method invocations and field accesses that a
developer will use at a given invocation site. In languages such as
Python, Java, and C\# an API member of a receiver object or namespace \code{foo}
is accessed by using a single dot. For example, a developer writes
\code{np.array} to access the \code{array} API from the \code{numpy} package.
IDEs commonly offer a list of candidate suggestions when a developer
presses the ``\code{.}'' key.
Although this is just one form of code completion it is one of the
most valuable and hardest~\citep{hellendoorn2019code}; this section
focuses on it. Note that our framework and models are more
general and could be used in other settings; wherever a candidate provider 
\candidateProvider for a completion context can be coded, our approach will
be applicable. In this section, we first explore how various architectural decisions
affect API completion models. Then, we contrast our models
to those of other publications, and discuss how our models
generalize to new APIs.

To evaluate the API completion performance, we follow the established evaluation
methodology within this area~\citep{karampatsis2019maybe}:
we assume that code is written sequentially and from left to right and measure if
the models predict the method that the user intended to invoke, directly reflecting
the use case of code completion systems. It is known that such an ``offline'' evaluation
does not fully reflect real-life (``online'') code completion setting~\citep{hellendoorn2019code}.
However, thanks to extensive data collection within Facebook,
\citet{aye2020learning} empirically observe a strong correlation 
between online and offline settings: 
although the accuracy of online completions is lower than those predicted offline,
models that perform better offline, also perform better in the online setting.
This is an encouraging result for research where due to privacy and confidentiality
it is impossible to monitor code completions of realistic and diverse
developer environments at scale across different organizations.
We also deploy our code completion system to a small set of about 2k users
and reconfirm these observations.

\boldpara{Data}
The training, validation and test data used for the experiments came from the 2700
top-starred Python source code repositories on GitHub. Scraped datasets commonly contain
a large amount of duplicates~\citep{lopes2017dejavu,allamanis2019adverse}, which
may skew the evaluation results. To avoid this, we deduplicate our dataset using the tool of \citet{allamanis2019adverse}.
Our dataset contains libraries from a diverse domains including scientific computing,
machine learning, dataflow programming, and web development.
To prepare the data, we use the Visual Studio Code type inference engine
(\href{https://github.com/Microsoft/PTVS}{PTVS})
to collect all completion locations where an API completion suggestion would be emitted
by PTVS, similar to \citet{svyatkovskiy2019pythia}. We extract the previous $N=80$ tokens preceding the method invocation ($\vect{t}_{\text{cx}}$),
and information about the receiver object or namespace,
which we use to create a Python \staticAnalysisProvider candidate completion provider.
The final dataset contains 255k files and a total
of 7.4 million API completion instances.
We split the data per-file into training-validation-test
at 60-20-20 proportions.
The list of GitHub URLs of the projects used
can be found at \url{https://gist.github.com/mallamanis/ce633f79dddb12bd8cb915704bcaace3}.

\boldpara{Evaluation Metrics}
We measure three aspects: accuracy,
model size and suggestion speed that directly relate to the
user experience of code completion.
To measure \textsl{model size}, we
compute the total number of parameters of each neural model. This number
correlates well with the RAM consumption
across machines and is not affected by noise (\eg garbage collection).
Given that every parameter is a \code{float32} we can compute the
(uncompressed) size of the parameters of each neural model
 similar to \citet{proksch2015intelligent}.

To measure the \textsl{computational cost} per-suggestion, we
compute the average time needed for our neural network to compute
a single suggestion on a CPU. Although we train our models
on a GPU we cannot expect that the developer environment has GPU hardware
and therefore only CPU time is relevant.
Furthermore, to match a realistic running environment,
we do \emph{not} batch the computation of suggestions when calculating
these statistics. Note here that the measurement is performed directly
with the PyTorch code. In practice, the neural network computation
would be statically compiled (\eg via ONNX, TorchScript, TFX). Again,
we expect such methods to yield similar improvements across configurations and do not
test them. Note that this time excludes any computation needed for a
static analysis, which is orthogonal to our models and is commonly
amortized across editing time. The experience and user studies of the Visual Studio
product team suggests that the computational budget should be at most a few tens
of milliseconds even on relatively old machines~\citep{svyatkovskiy2019pythia}.

Finally, we are interested in evaluating the \textsl{predictive accuracy} of each model,
\ie how well each completion model ranks the intended candidate higher.
This directly relates to the user experience and the usefulness of any code completion
system: the higher a relevant result is ranked, the better the accuracy of the model.
We use two well established metrics. First, ``recall at top $k$''\footnote{The
terminology derives from information retrieval domain. Some
papers refer to this metric as ``accuracy at top $k$'' instead.} (denoted as ``Recall@$k$'')
measures the proportion of examples where the correct completion
is in the top $k$ suggestions. We report $k=1$ and $k=5$ as we do not
expect users to look at the suggestions beyond that point~\citep{miller1956magical}.
We also report the mean reciprocal rank (MRR), which is commonly used for evaluating
ranking methods. MRR takes values in $[0,1]$, with 1 being the best
score, and is defined as $\frac{1}{N}\sum_{i=1}^N \frac{1}{r_i}$
where $r_i$ is the rank of each target suggestion.

Each of our experiments runs on a virtual machine equipped with one NVIDIA Tesla V100 GPU
and an Intel Xeon E5-2690 v4 (Broadwell) CPU. The results
presented in this section stem from more than 1 year of a single GPU-time
(across multiple machines).
It should be noted that various techniques, such as model quantization~\citep{courbariaux2014training},
can be applied. However, these methods commonly
provide similar speed-ups and memory reductions across
all models retaining their relative ordering.
We thus ignore these techniques at this stage.

\subsection{Multitarget Evaluation}
\label{subsec:multiobjective eval}
Although predictive performance is an important factor, it is important to
consider the memory and computational cost trade-offs to offer the best
experience to developers. These factors have been mostly overlooked
by the current literature. Improving on the last two quantities
usually reduces predictive accuracy. Therefore, we treat
the evaluation as multitarget.

\begin{figure*}[tp]
	\centering
	\begin{subfigure}[b]{\textwidth}
		\includegraphics[width=\textwidth]{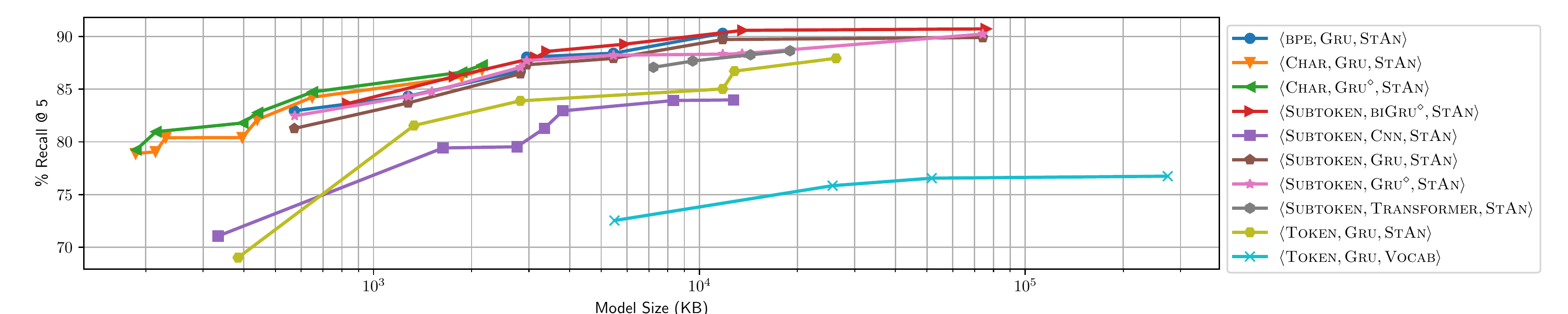}
		\caption{Recall@5 ($\uparrow$) \vs Model Size ($\leftarrow$). Note semi-log$x$ scale.}\label{fig:acc vs size}
		\vspace{0.5em}
	\end{subfigure}

	\begin{subfigure}[b]{\textwidth}
		\includegraphics[width=\textwidth]{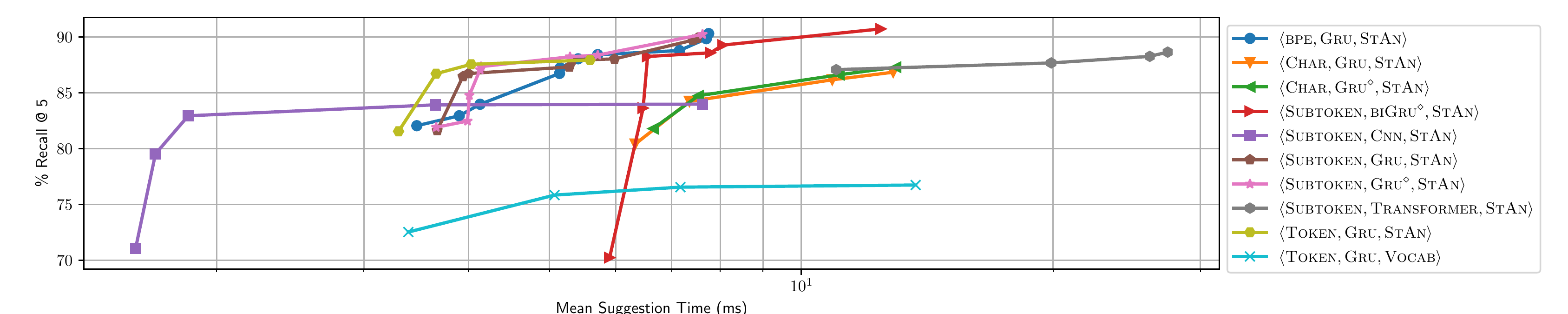}
		\caption{Recall@5 ($\uparrow$) \vs Mean Suggestion Time ($\leftarrow$). Note semi-log$x$ scale.}\label{fig:acc vs time}
		\vspace{0.5em}
	\end{subfigure}

	\begin{subfigure}[b]{\textwidth}
		\includegraphics[width=\textwidth]{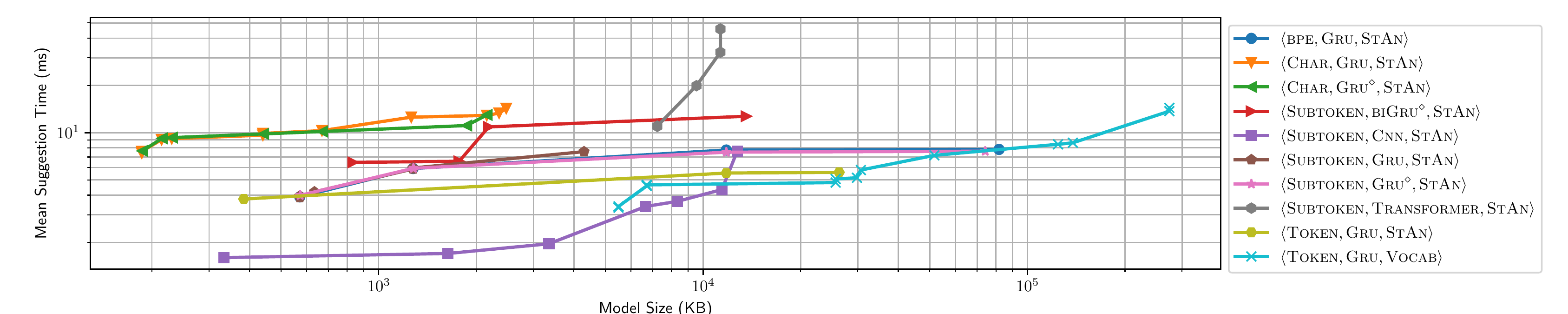}
		\caption{Mean Suggestion Time ($\downarrow$) \vs Model Size ($\leftarrow$). Note log-log scale.}\label{fig:time vs size}
	\end{subfigure}
	\caption{Pareto fronts (Recall@5 \vs mean suggestion time \vs model size) across some model configurations
	 over our hyperparameter search. Note some axes are in log-scale.
	Arrows ($\uparrow, \downarrow, \leftarrow, \rightarrow$) denote which direction
	improves the given axis. When a front stops early, it means that the hyperparameter search did not
	find settings that were Pareto optimal beyond that point. We do \emph{not} plot the Pareto fronts of 
	\vocabProvider models since they are so much worse that would require increasing the scale
	of the plot.}\label{fig:Paretos}
\end{figure*}

To achieve this, we run a search across multiple model
configurations and hyperparameters\footnote{We use an automatic
hyperparameter tuning method to perform a random search over 
ranges of all hyperparameters. For some hyperparameters, such as learning
rate, the search is uniform random at a log scale.}.
These --- among other parameters --- include the size of the vocabularies
($V_t$ for \tokenUnitEncoder and $V_s$ for \subtokenEncoder), the context encoder
hidden dimension $H$, and the size $D$ of the token embeddings
which have the biggest effect on a model's size. Reporting the results for
all configurations would significantly reduce the clarity of our evaluation.
Therefore, we gradually bisect the design space and discuss
the results. \autoref{fig:Paretos} plots
the Pareto fronts of the multitarget evaluation across our evaluation metrics for some of the model configurations,
as computed by our search. Each line represents the Pareto optimal options
for a given configuration.
\autoref{tbl:detailed results} shows the evaluation metrics for a selected
subset of model configurations. For each configuration, we present the metrics of
the best models that are approximately 3 MB and 50 MB in size.
We pick these sizes because they
represent two realistic points. A 3 MB model can be deployed even in
severely restricted environments (\eg low-end hardware used to teach
students to code). A 50 MB model is a reasonable upper bound
size for a plugin in a modern IDE or editor. Note that although most modern
computers have a few gigabytes of RAM, modern IDEs --- such as Visual Studio Code ---
contain a number of components which individually amount to a few megabytes of RAM
but collectively can consume a significant proportion of a computer's RAM.
Although it may be surprising to the reader, the experience of the Visual Studio teams
deploying IDEs to thousands of users, suggests that the memory consumption
should be as minimal as possible and --- ideally --- less than 50 MB.

\begin{table*}[tp]
	\centering
	\caption{Detailed evaluation for a selected subset of model configurations. We show results for the best performing
		model closest to two model sizes. Computational time ranges denote standard deviation.
		For some configurations, there is no model of $\sim$ 50 MB that outperforms a model of $\sim$ 3 MB.
		This is denoted with a dash (---).} \label{tbl:detailed results}
	\begin{tabular}{lll rrrrrrrrr} \toprule
		\multirow{2}{*}{\tokenEncoder}   &   \multirow{2}{*}{\contextEncoder}      &  \multirow{2}{*}{\candidateProvider} &
																		  \multicolumn{4}{c}{Best for size $\sim$3 MB} && \multicolumn{4}{c}{Best for size $\sim$50 MB} \\ \cmidrule{4-7} \cmidrule{9-12}
						&                        &                          & Recall@1 & Recall@5 &  MRR & Time (ms)                       && Recall@1 & Recall@5 & MRR    &  Time (ms)  \\ \midrule
	  \multicolumn{3}{l}{Popularity (Most Frequently Used)}                & 41.75    &   72.04  & 0.5470 & 0.02 $\pm$ 0.01               &&  ---     &  ---     &  ---   &  ---         \\
	  \tokenUnitEncoder &   \rnnCxEncoder        &  \vocabProvider         & 53.01    &   72.53  & 0.6140 & 3.39 $\pm$ 1.00               &&   55.87  &   76.55  & 0.6477 &  7.17 $\pm$ 1.43 \\
	  \tokenUnitEncoder &  \transformerCxEncoder &  \vocabProvider         & 24.16    &   40.52  & 0.3103 &10.46 $\pm$ 3.00               &&   55.48  &   74.26  & 0.6354 & 36.73 $\pm$ 5.01 \\
	  \tokenUnitEncoder &   \rnnCxEncoder        &  \staticAnalysisProvider& 63.78    &   83.89  & 0.7245 & 3.71 $\pm$ 0.98               &&   68.78  &   87.93  & 0.7703 &  5.59 $\pm$ 1.28 \\
	  \subtokenEncoder  &   \rnnCxEncoder        &  \staticAnalysisProvider& 63.40    &   87.31  & 0.7369 & 5.28 $\pm$ 1.13               &&   67.98  &   89.90  & 0.7744 &  7.51 $\pm$ 1.78 \\
	  \bpeEncoder       &   \rnnCxEncoder        &  \staticAnalysisProvider& \textbf{66.35}&   88.04  & \textbf{0.7567} & 5.41 $\pm$ 1.50 && \textbf{70.09}&   89.84  &\textbf{0.7861}&  7.70 $\pm$ 1.85 \\
	  \charEncoder      &   \rnnCxEncoder        &  \staticAnalysisProvider& 64.30    &   86.84  & 0.7396 &12.88 $\pm$ 3.25               &&  ---     &  ---     &  ---   &  ---            \\
	  \subtokenEncoder  &\rnnCxEncoder$^\diamond$&  \staticAnalysisProvider& 63.76    &   87.71  & 0.7408 & 5.40 $\pm$ 1.23               &&   66.57  &   90.23  & 0.7681 &  7.63 $\pm$ 1.97 \\
	  \subtokenEncoder  &\birnnCxEncoder$^\diamond$&\staticAnalysisProvider& 64.25    &   \textbf{88.58}  & 0.7428 & 7.79 $\pm$ 1.37      &&   67.17  &\textbf{90.58}& 0.7736 & 12.72 $\pm$ 2.42\\
	  \subtokenEncoder  &   \cnnCxEncoder        &  \staticAnalysisProvider& 55.66    &   81.29  & 0.6652 & 1.96 $\pm$ 0.89               &&   57.19  &   83.98  & 0.6834 &  7.62 $\pm$ 1.77 \\
	  \subtokenEncoder  &\transformerCxEncoder   &  \staticAnalysisProvider& 61.81    &   87.07  & 0.7241 & 11.01$\pm$ 2.18               &&   65.36  &   87.36  & 0.7504 &  25.85$\pm$ 3.91  \\
	\bottomrule
	\end{tabular}
	\vspace{0.5em}
\end{table*}


We additionally compare our models to a simple baseline (``Popularity'') that yields
the most frequently used target completion for a given API in our training set.
This can be thought as a non-neural
\modelName{\tokenUnitEncoder}{\emptyset}{\staticAnalysisProvider} model.
The predictive performance of the baseline is worse than all neural models,
although it is faster and smaller in size (\autoref{tbl:detailed results}).
This is unsurprising, since it does \emph{not} take into
account any context information beyond what is available
to the \staticAnalysisProvider provider.

\boldpara{\staticAnalysisProvider \vs \vocabProvider}
\autoref{tbl:detailed results} shows that \uline{\vocabProvider-based models
underperform significantly compared to our novel \staticAnalysisProvider-based models.} The \vocabProvider models are
similar to the language models presented by \citet{hellendoorn2017deep}
except that our models are trained only on API completion locations,
and not for predicting arbitrary code tokens.
We do \emph{not} plot the Pareto fronts of \vocabProvider models in \autoref{fig:acc vs size} since
they are so much worse that would require increasing the scale
of the plot. This is not surprising since
\staticAnalysisProvider models have strictly more information.
All \vocabProvider-based models have multiple shortcomings.
First, they need to predict the target completion from a long list of
plausible candidates, without any explicit knowledge of the correct choices,
and thus have multiple opportunities for making errors: making a correct
choice from a longer list is harder than from a short list. 
Moreover, \vocabProvider-based models are larger and slower.
For a \vocabProvider-based model to have good recall, its
vocabulary must be sufficiently large to contain the majority of suggestions
it may need to make. In turn, a large vocabulary
implies a large embedding matrix, which substantially increases
memory requirements. The relatively slow speed of this model stems from
the need to compute \autoref{eqn:prob} over the whole vocabulary $V_t$, whereas
\staticAnalysisProvider-based models can compute it over only a
smaller set of candidate completions.
These observations are common across all \vocabProvider models, even
those not explicitly presented here. For this reason we will not consider
them any further.

\boldpara{Token Encoders \tokenEncoder}
Having established that \staticAnalysisProvider-based models are preferable to \vocabProvider-based
models, we now discuss the different token encoders using the \staticAnalysisProvider candidate provider
and the \rnnCxEncoder context encoder. The performance differences are small,
but generally the \tokenUnitEncoder models perform worse. The difference is
more pronounced for models with a smaller vocabulary, which fail to
represent the sparsity of code tokens. In \autoref{subsec:generalization} we
show that \tokenUnitEncoder-level encoders also tend to generalize worse compared
to other token encoders. \uline{The \subtokenEncoder, \bpeEncoder, and \charEncoder provide
competitive results at a higher computational cost} (needed for composing the
representation of each token). \charEncoder models perform best for small model
sizes ($\le$1 MB), however with increased model capacity, they fail to scale up.
In light of these results, we consider \subtokenEncoder or \bpeEncoder based models to be
reasonable options. These observations about \bpeEncoder are consistent with those
of \citet{karampatsis2019maybe} and \citet{aye2020sequence}.

\boldpara{Context Encoders \contextEncoder}
Finally, we turn our attention to the context encoders. We select
the \subtokenEncoder encoder as it provides a reasonable trade-off between memory
consumption and predictive performance. Both \cnnCxEncoder and \transformerCxEncoder
underperform compared to \rnnCxEncoder encoders. Additionally, \transformerCxEncoder-based
models are significantly more computationally expensive.
\birnnCxEncoder encoders improve marginally over \rnnCxEncoder, but at an increased
computational cost.
Finally, completion-location annotated ($\diamond$) context encoders provide a small
but consistent improvement by using the additional information.

\boldpara{Model Size \vs Computational Cost}
\autoref{fig:time vs size} shows the log-log Pareto fronts between the
computational time needed for computing a single suggestion \vs
the model size. As a general principle, the larger a model (\ie
more parameters it contains) the slower the computation of the
completion suggestions. Furthermore, different configurations
have different scaling behaviors. \transformerCxEncoder-based
models are --- unsurprisingly --- the slowest, whereas the model size of other models
seems to have a smaller but noticeable effect.
\charEncoder-based models tend to be slower compared to \subtokenEncoder or
\tokenUnitEncoder-based models, since they trade-off memory with computation.
Overall, \uline{most models make predictions in under 20 ms which
makes them eligible for real-time code completion systems}.

\boldpara{Comparison with Models from Other Publications}
We select the best 3MB and 50MB \modelName{\bpeEncoder}{\rnnCxEncoder}{\staticAnalysisProvider}
as our models of choice which achieve 66\% (resp. 70\%) recall@1 with about
6ms (resp. 8ms) of average suggestion time.
Direct comparison with models discussed in other publications in
terms of predictive accuracy is not possible due to
differences in datasets and programming languages; none of
the pre-existing datasets can be used since we cannot run
the necessary static analyses on them (for the \staticAnalysisProvider provider)
and often the datasets are not available.
Our emphasis here is to show the qualitative differences and
similarities with prior work.
Note also that \citet{karampatsis2019maybe}
train on different code corpora, some of which are known to contain
duplicates~\citep{allamanis2019adverse} hence further making any direct
comparison impossible. 

Nevertheless, we can
reason from first principles to discuss the advantages of \staticAnalysisProvider-based
models and our framework over existing work, while retaining the
advances from the methods of \citet{karampatsis2019maybe} and \citet{aye2020sequence}.
First, \citet{karampatsis2019maybe} definitively show that $n$-gram
language models have worse predictive performance than neural
models due to their better generalization over unseen (sub)tokens.
More importantly, reasonably well-performing $n$-gram models tend to
have unacceptably large sizes that are between 5-8.5GB~\citep{karampatsis2019maybe}
($\times$100 to $\times$1700 more than our models).
The BPE-based neural language model of \citet{karampatsis2019maybe} has a size
of 115MB which is about $\times$2--$\times$38 larger than our models, while
similar to our \modelName{\bpeEncoder}{\rnnCxEncoder}{\vocabProvider} models.
\citet{aye2020sequence} train a language model similar to \citet{karampatsis2019maybe}
with a size of 402.6~MB ($\times$8--$\times$134 larger than ours).

As discussed earlier, predictive performance comparison with models
with structural prediction providers (\eg the beam
search over BPE tokens of \citet{aye2020sequence,karampatsis2019maybe}) would be pointless.
Reasoning from first principles, \staticAnalysisProvider
models have access to strictly more information (from the static 
analysis) and need to ``just'' pick
suggestion from a short list of candidates. In contrast,
structural providers
need to predict a valid API token out of an \emph{unbounded} vocabulary
of subtokens. Thus, by construction, \uline{\staticAnalysisProvider-based
models have an advantage (access to more information) and will always
perform better to similar non-\staticAnalysisProvider models.}
Essentially, \staticAnalysisProvider providers allows us to
maintain all the novel advances of BPE-based language models of \citet{aye2020sequence,karampatsis2019maybe}
without any of the disadvantages of the structural prediction.
Furthermore, since the target task is simpler in \staticAnalysisProvider-based
models, models with similar predictive performance can be smaller --- and thus faster.
Finally, structural completion models perform a computationally
costly beam search, which further increases the computational cost.
However, if the static analysis backing \staticAnalysisProvider
is unsound, erroneous, or uses heuristics that hurt precision/recall, then
these observations will not hold.

\subsection{Deployment}
We deploy our 50MB $(\subtokenEncoder, \rnnCxEncoder, \staticAnalysisProvider)$
completion model to Visual Studio and Visual Studio Code users. The
model is converted into ONNX, and quantized. Due to
strict privacy and confidentiality concerns we \emph{cannot} observe any specifics
about the code completions (\eg the code context) but
we can measure the performance of the suggestions served.
We observe $\gg$100k code completion events per workday for many months, with a mean serving time
of 20 ms (end-to-end; including the static analysis and UI rendering).
Similarly to \citet{aye2020learning},
we observe a reduced predictive performance compared to the one
measured in the offline experiments with Recall@5 of about 70\% 
(\vs 90.23\% in the offline experiments). We posit that this is due
to differences between the open-source code used during training and the
code used seen in practice, along with the differences of partial code
to commited code found on GitHub. Our model substantially improves upon the
previous non-neural model deployed~\citep{svyatkovskiy2019pythia} both in terms of
predictive accuracy (+50\%) and serving speed.

\subsection{Generalization to Unseen APIs} \label{subsec:generalization}
\begin{table}\centering
	\caption{MRR on unseen libraries for models with size $\sim$50 MB with a \rnnCxEncoder context encoder.}\label{tbl:generalization experiments}
	\begin{tabular}{llrrrrr} \toprule
		\tokenEncoder      &  \candidateProvider & \href{https://github.com/google/jax}{\code{jax}} &&
				  \href{https://github.com/horovod/horovod}{\code{horovod}} &&
				  \href{https://github.com/apache/spark/tree/master/python}{\code{pyspark}}  \\ \midrule
\tokenUnitEncoder & \vocabProvider          &  0.0931           && 0.2526  && 0.0121 \\
\subtokenEncoder   & \staticAnalysisProvider &  0.4184          && \textbf{0.6051}  && 0.5410 \\
\bpeEncoder        & \staticAnalysisProvider &\textbf{0.5444}   && 0.5430  && \textbf{0.5574} \\
 \midrule
Random& \staticAnalysisProvider &  0.2512           && 0.2628  && 0.4901 \\
	\bottomrule \end{tabular}
\end{table}

So far, we observed the performance of the code completion models when
the target completion APIs were seen during training. However,
APIs evolve and code completion systems are asked to complete code from
previously unseen libraries or new user-defined code. In all these cases,
we cannot expect to have training data to train accurate models. Instead, we
hope that models generalize.

To test these scenarios, we held out from our dataset API completions for
three libraries. \autoref{tbl:generalization experiments} shows
the results of how the neural completion models generalize to completions of the unseen libraries.
Our baseline (bottom line) is a completion system that randomly ranks
the candidate code completions. This is a weaker baseline than
the ``most frequently used'' baseline (in \autoref{subsec:multiobjective eval})
but is the only reasonable choice for this scenario, since we have no
prior information about the APIs.
\autoref{tbl:generalization experiments} shows that all models, except the
\vocabProvider-based ones, perform better than the baselines, which
indicates that all neural models generalize to some extent.
The bad performance of the \vocabProvider model is expected as it
has to generate the names of the target completion candidates from its
vocabulary. Many of those names have \emph{not} been previously
seen and therefore these models fail to generalize. Additionally, the \tokenUnitEncoder-based
model performs consistently poor. This can be attributed to the
fact that \tokenUnitEncoder-based models cannot generalize easily
to previously unseen tokens. In contrast, the models that encode
tokens in a more granular way tend to perform better, although
there is not a clear ``winner''. The observed differences may be
attributed to different naming conventions of the tested libraries.
The results presented here suggest that our novel \uline{\staticAnalysisProvider-based
models with granular \bpeEncoder token encodings 
are also preferable from a generalization perspective}.

\subsection{Training without a Static Analyzer}

To train a \staticAnalysisProvider-based model, we needed
to build a dataset making sure that a static analysis tool is able
to resolve the developer environment (\eg libraries, other dependencies).
However, getting a static analyzer to run for a sufficiently large codebase
is hard and does not easily scale, a known issue in the area of ``big code''~\citep{allamanis2018survey}.
Inspired by techniques such as NCE~\citep{gutmann2012noise},
we hypothesize that we may be able to overcome this challenge by 
using a proxy candidate provider that yields random distractor candidate completion targets. (\ie negative samples).
Furthermore, for training efficiency, we can use
the target completions of other samples in each minibatch as distractors.



We test this on the setting of the best \modelName{\bpeEncoder}{\rnnCxEncoder}{\staticAnalysisProvider}
configuration of approximately 50 MB. The results show some
degradation in performance. For example, Recall@5 falls
to 85.5\% (from 89.9\%) and MRR falls to 0.729 (from 0.786).
This disproves our hypothesis that a proxy static analyzer
can be used. This degradation can be
attributed to the fact that model capacity is spent for
learning to avoid distractor candidate target completions that
will never appear at test time.

There is a threat to the validity to this observation. Our
testset contains samples where we could use --- at batch --- a
static analyzer to retrieve candidate completions. Thus, our
testset is slightly biased towards locations where a
\staticAnalysisProvider could be used and
provides no indication about the performance at completion locations
where a \staticAnalysisProvider needs manual configuration. However, our
testset contains diverse APIs, which gives sufficient confidence
that our results hold more generally.

\boldpara{Removing the Explicit Subtoken Vocabulary}
So far, we have measured the size of a model by counting the parameters
of the neural model. However, there is an additional cost for all
non-\charEncoder models. We need to store in memory a mapping from
the string representations of tokens in the vocabulary to their unique ids.
This mapping (commonly implemented as a hash map) is consulted when performing an embedding lookup
(\eg in the $\textsc{EmbeddingLookUp}(t, V_t)$ of \autoref{eq:tokenEncoder} and \autoref{eq:subtokenEncoder})
and maps each (sub)token into a unique index in the embedding matrix.
However, this data structure consumes memory. For example, a
\subtokenEncoder-model with a vocabulary of 10k subtokens has a hash map that consumes about 1.1 MB of RAM.
This additional cost of storing the necessary metadata \emph{cannot} be avoided on
\bpeEncoder-based models where this metadata is necessary or in \vocabProvider-based
models where the target completion needs to be generated.
However, for  non-\bpeEncoder \staticAnalysisProvider models
we can employ feature hashing to eliminate the hash map.

Feature hashing refers to a common trick for vectorizing features~\citep{attenberg2009feature,moody1989fast}
and is a form of dimensionality reduction. The core idea
is that a good hash function $\phi$ can provide a reasonable, deterministic mapping
of features to ids. As with all hashing methods, this introduces
collisions as multiple features may have the same hash, which nevertheless
the machine learning models can learn to overcome to some extent. We use this technique
for subtokens in the \modelName{\subtokenEncoder}{\rnnCxEncoder}{\staticAnalysisProvider}
model. For each subtoken, we compute the MD5 hash of its string representation and map
it to an integer in
$\{0 ... |V|-1\}$ where $|V|$ is the size of the ``vocabulary'', \ie $\phi(s) = \textsc{Md5}(s) \mod |V|$. The larger $|V|$,
the fewer hash collisions at the cost of a larger embedding matrix.
Experimentally, we observe minor differences compared to the original models. Specifically,
a model with $|V|=2500$, sees a reduction of MRR of about 1\%,
whereas the performance of models with larger $|V|$ remains unchanged. Altogether,
the results suggest that we can further reduce the memory consumption of
\subtokenEncoder models by eliminating the stored hash map,
without any impact on predictive performance.

\section{Related Work}
\label{sect:related_work}
Our work is related to a large set of literature in natural language processing and machine learning,
particularly deep learning. We refer the interested reader to the book of \citet{goodfellow2016deep}
for a detailed treatise of deep learning methods and focus the rest of this section on code
completion.

The first to propose learning code completions from data were arguably \citet{bruch2009learning},
who tested three methods: association rule mining, frequency-based models
and nearest-neighbor-based methods. That stream of work evolved into the
Eclipse Code Recommenders~\citep{eclipseCodeRecommenders} and was --- to our
knowledge --- the first data-driven code completion system to be deployed widely 
to a popular IDE. However, as of 2019 this project
has been discontinued~\citep{eclipseCodeRecommenders}. Following similar principles, \citet{proksch2015intelligent} presented a
Bayesian network for code completion. Similar to our \staticAnalysisProvider-based models, both these works
focus on code completion within specific contexts, \eg when the developer creates a method invocation.
However, in contrast to our work, the methods of \citet{bruch2009learning} and \citet{proksch2015intelligent} rely
heavily on manually extracting features that are relevant to the completion task. For example,
\citet{proksch2015intelligent} extract features such as the direct supertype of the enclosing
class, the kind of the definition, \etc. In contrast, our models
require minimal manual feature extraction and can instead employ information that is
readily available using a compiler (lexemes, variable bindings, \etc).

Our work is centered around the area of machine learning models for source code~\citep{allamanis2018survey}.
The core principle is to avoid extracting hand-coded features and instead rely on (possibly) structured representation
of code (lexemes, ASTs, dataflow, \etc) and learn directly a task over those representations.
``Feature-less'' machine learning models of code completion were first studied by
\citet{hindle2012naturalness} who create a token-level $n$-gram language model completion
models.  This was arguably the first
model that performed unconstrained code completion.
A variety of language models have since been explored~\citep{allamanis2013mining,nguyen2013statistical,bielik2016phog}.
Further improvements to language models (and therefore completion systems)
were made by \citet{tu2014localness} who noticed that source code tends to have a localness
property, \ie tokens tend to be repeated locally. The authors showed that by introducing a cache-based
language model~\citep{kuhn1990cache}, performance could be improved. \citet{hellendoorn2017deep}
then set to test neural models of code along with $n$-gram language models for the
task. Their evaluation showed that carefully tuning an $n$-gram language model, with a
hierarchically scoped cache, outperforms some neural models. \citet{nguyen2019combining}
used static analyses to extract features and build a templated $n$-gram language model
for code completion which nevertheless cannot learn generalize to previous unseen APIs. To our knowledge,
none of the $n$-gram language models has been deployed in a real-life IDE, due to the
size constraints.

Deep learning methods are central to ``feature-less'' models and eliminate the need for hand-coded features
across domains (\eg image recognition) and can directly learn from raw data.
Within this context, a multitude of deep learning models have been researched for source code.
Recently, \citet{karampatsis2019maybe} showed that an appropriately designed neural model that uses byte-pair
encoding (BPE) yields superior results to non-deep learning models including those of \citet{hellendoorn2017deep}.
We replicated the advantages of BPE in this work.  Despite this, the model of
\citet{karampatsis2019maybe} treats completion as a generation
problem over BPE subtokens instead of taking advantage of readily available information from
static analyses, which simplifies the task and offers improved results.
Simultaneously, \citet{svyatkovskiy2019pythia} presented a neural model that corresponds to our
 \modelName{\tokenUnitEncoder}{\rnnCxEncoder}{\vocabProvider} model, which our 
 \modelName{\subtokenEncoder}{\rnnCxEncoder}{\staticAnalysisProvider} model outperform.
Although other, more structured, language models of code have been researched~\citep{maddison2014structured,bielik2016phog,brockschmidt2018generative},
none of those have been tested for practical code completion systems. This is because of
the complexity of the used code representations: since completion contexts are
commonly incomplete (\eg they do not parse), sophisticated methods are required to extract the structured representations
needed by these models.

\section{Discussion \& Open Challenges}
\label{sect:discussion}
We presented an exploration of the
design space of practical neural code completion and showed
how to combine different deep learning components
to retrieve a range of trade-offs beyond reusing existing neural
architectures. Within this framework,
we implemented and evaluated a number of neural code completion
models aiming to improve their performance characteristics. The
subtoken-level static analysis-based models strike the best trade-off among
predictive performance, model size and computational speed
across the tested models. Such models are being deployed in Visual Studio
with an acceptable memory and computation footprint.
This is an important step towards providing inclusive tools
to all developers, even those that cannot access state-of-the-art equipment.


\section*{Acknowledgement}
We thank the Visual Studio Data Science
team for their help and Marc Brockschmidt for
useful discussions.

\bibliographystyle{IEEEtranN}
\bibliography{bibliography}

\begin{thebibliography}{46}
\providecommand{\natexlab}[1]{#1}
\providecommand{\url}[1]{#1}
\csname url@samestyle\endcsname
\providecommand{\newblock}{\relax}
\providecommand{\bibinfo}[2]{#2}
\providecommand{\BIBentrySTDinterwordspacing}{\spaceskip=0pt\relax}
\providecommand{\BIBentryALTinterwordstretchfactor}{4}
\providecommand{\BIBentryALTinterwordspacing}{\spaceskip=\fontdimen2\font plus
\BIBentryALTinterwordstretchfactor\fontdimen3\font minus
  \fontdimen4\font\relax}
\providecommand{\BIBforeignlanguage}[2]{{%
\expandafter\ifx\csname l@#1\endcsname\relax
\typeout{** WARNING: IEEEtranN.bst: No hyphenation pattern has been}%
\typeout{** loaded for the language `#1'. Using the pattern for}%
\typeout{** the default language instead.}%
\else
\language=\csname l@#1\endcsname
\fi
#2}}
\providecommand{\BIBdecl}{\relax}
\BIBdecl

\bibitem[Allamanis et~al.(2018{\natexlab{a}})Allamanis, Barr, Devanbu, and
  Sutton]{allamanis2018survey}
M.~Allamanis, E.~T. Barr, P.~Devanbu, and C.~Sutton, ``A survey of machine
  learning for big code and naturalness,'' \emph{ACM Computing Surveys (CSUR)},
  vol.~51, no.~4, p.~81, 2018.

\bibitem[Bruch et~al.(2009)Bruch, Monperrus, and Mezini]{bruch2009learning}
M.~Bruch, M.~Monperrus, and M.~Mezini, ``Learning from examples to improve code
  completion systems,'' in \emph{Proceedings of the Joint Meeting of the
  European Software Engineering Conference and the Symposium on the Foundations
  of Software Engineering (ESEC/FSE)}, 2009.

\bibitem[Hindle et~al.(2012)Hindle, Barr, Su, Gabel, and
  Devanbu]{hindle2012naturalness}
A.~Hindle, E.~T. Barr, Z.~Su, M.~Gabel, and P.~Devanbu, ``On the naturalness of
  software,'' in \emph{Proceedings of the International Conference on Software
  Engineering (ICSE)}, 2012.

\bibitem[Nguyen et~al.(2013)Nguyen, Nguyen, Nguyen, and
  Nguyen]{nguyen2013statistical}
T.~T. Nguyen, A.~T. Nguyen, H.~A. Nguyen, and T.~N. Nguyen, ``A statistical
  semantic language model for source code,'' in \emph{Proceedings of the Joint
  Meeting of the European Software Engineering Conference and the Symposium on
  the Foundations of Software Engineering (ESEC/FSE)}, 2013.

\bibitem[Amann et~al.(2016)Amann, Proksch, Nadi, and Mezini]{amann2016study}
S.~Amann, S.~Proksch, S.~Nadi, and M.~Mezini, ``A study of {V}isual {S}tudio
  usage in practice,'' in \emph{Proceedings of the International Conference on
  Software Analysis, Evolution, and Reengineering (SANER)}, 2016.

\bibitem[Murphy et~al.(2006)Murphy, Kersten, and Findlater]{murphy2006java}
G.~C. Murphy, M.~Kersten, and L.~Findlater, ``How are {J}ava software
  developers using the {E}clipse {IDE}?'' \emph{IEEE software}, vol.~23, no.~4,
  pp. 76--83, 2006.

\bibitem[Proksch et~al.(2015)Proksch, Lerch, and
  Mezini]{proksch2015intelligent}
S.~Proksch, J.~Lerch, and M.~Mezini, ``Intelligent code completion with
  {B}ayesian networks,'' \emph{ACM Transactions on Software Engineering and
  Methodology (TOSEM)}, 2015.

\bibitem[Hellendoorn and Devanbu(2017)]{hellendoorn2017deep}
V.~J. Hellendoorn and P.~Devanbu, ``Are deep neural networks the best choice
  for modeling source code?'' in \emph{Proceedings of the International
  Symposium on Foundations of Software Engineering (FSE)}, 2017.

\bibitem[Karampatsis et~al.(2020)Karampatsis, Babii, Robbes, Sutton, and
  Janes]{karampatsis2019maybe}
R.-M. Karampatsis, H.~Babii, R.~Robbes, C.~Sutton, and A.~Janes, ``Big code!=
  big vocabulary: Open-vocabulary models for source code,'' in
  \emph{Proceedings of the International Conference on Software Engineering
  (ICSE)}, 2020.

\bibitem[Svyatkovskiy et~al.(2019)Svyatkovskiy, Zhao, Fu, and
  Sundaresan]{svyatkovskiy2019pythia}
A.~Svyatkovskiy, Y.~Zhao, S.~Fu, and N.~Sundaresan, ``Pythia: {AI}-assisted
  code completion system,'' in \emph{Proceedings of the 25th ACM SIGKDD
  International Conference on Knowledge Discovery \& Data Mining}, 2019, pp.
  2727--2735.

\bibitem[Aye and Kaiser(2020)]{aye2020sequence}
G.~A. Aye and G.~E. Kaiser, ``Sequence model design for code completion in the
  modern {IDE},'' \emph{arXiv preprint arXiv:2004.05249}, 2020.

\bibitem[Allamanis and Sutton(2013)]{allamanis2013mining}
M.~Allamanis and C.~Sutton, ``Mining source code repositories at massive scale
  using language modeling,'' in \emph{Proceedings of the Working Conference on
  Mining Software Repositories (MSR)}, 2013.

\bibitem[Tu et~al.(2014)Tu, Su, and Devanbu]{tu2014localness}
Z.~Tu, Z.~Su, and P.~Devanbu, ``On the localness of software,'' in
  \emph{Proceedings of the International Symposium on Foundations of Software
  Engineering (FSE)}, 2014.

\bibitem[Bielik et~al.(2016)Bielik, Raychev, and Vechev]{bielik2016phog}
P.~Bielik, V.~Raychev, and M.~Vechev, ``{PHOG}: Probabilistic model for code,''
  in \emph{Proceedings of the International Conference on Machine Learning
  (ICML)}, 2016.

\bibitem[Maddison and Tarlow(2014)]{maddison2014structured}
C.~Maddison and D.~Tarlow, ``Structured generative models of natural source
  code,'' in \emph{Proceedings of the International Conference on Machine
  Learning (ICML)}, 2014.

\bibitem[Aye et~al.(2020)Aye, Kim, and Li]{aye2020learning}
G.~A. Aye, S.~Kim, and H.~Li, ``Learning autocompletion from real-world
  datasets,'' \emph{arXiv preprint arXiv:2011.04542}, 2020.

\bibitem[Allamanis et~al.(2018{\natexlab{b}})Allamanis, Brockschmidt, and
  Khademi]{allamanis2018learning}
M.~Allamanis, M.~Brockschmidt, and M.~Khademi, ``Learning to represent programs
  with graphs,'' in \emph{Proceedings of the International Conference on
  Learning Representations (ICLR)}, 2018.

\bibitem[Inan et~al.(2016)Inan, Khosravi, and Socher]{inan2016tying}
H.~Inan, K.~Khosravi, and R.~Socher, ``Tying word vectors and word classifiers:
  A loss framework for language modeling,'' \emph{arXiv preprint
  arXiv:1611.01462}, 2016.

\bibitem[Sennrich et~al.(2016)Sennrich, Haddow, and Birch]{sennrich2016neural}
R.~Sennrich, B.~Haddow, and A.~Birch, ``Neural machine translation of rare
  words with subword units,'' in \emph{Proceedings of the Annual Meeting of the
  Association for Computational Linguistics (ACL)}, 2016.

\bibitem[Zhang et~al.(2015)Zhang, Zhao, and LeCun]{zhang2015character}
X.~Zhang, J.~Zhao, and Y.~LeCun, ``Character-level convolutional networks for
  text classification,'' in \emph{Advances in neural information processing
  systems}, 2015, pp. 649--657.

\bibitem[Kim et~al.(2016)Kim, Jernite, Sontag, and Rush]{kim2016character}
Y.~Kim, Y.~Jernite, D.~Sontag, and A.~M. Rush, ``Character-aware neural
  language models,'' in \emph{Thirtieth AAAI Conference on Artificial
  Intelligence}, 2016.

\bibitem[Hochreiter and Schmidhuber(1997)]{hochreiter1997long}
S.~Hochreiter and J.~Schmidhuber, ``Long short-term memory,'' \emph{Neural
  Computation}, 1997.

\bibitem[Bahdanau et~al.(2015)Bahdanau, Cho, and Bengio]{bahdanau2014neural}
D.~Bahdanau, K.~Cho, and Y.~Bengio, ``Neural machine translation by jointly
  learning to align and translate,'' in \emph{Proceedings of the International
  Conference on Learning Representations (ICLR)}, 2015.

\bibitem[Kim(2014)]{kim2014convolutional}
Y.~Kim, ``Convolutional neural networks for sentence classification,''
  \emph{arXiv preprint arXiv:1408.5882}, 2014.

\bibitem[Vaswani et~al.(2017)Vaswani, Shazeer, Parmar, Uszkoreit, Jones, Gomez,
  Kaiser, and Polosukhin]{vaswani2017attention}
A.~Vaswani, N.~Shazeer, N.~Parmar, J.~Uszkoreit, L.~Jones, A.~N. Gomez,
  {\L}.~Kaiser, and I.~Polosukhin, ``Attention is all you need,'' in
  \emph{Advances in Neural Information Processing Systems}, 2017, pp.
  5998--6008.

\bibitem[Devlin et~al.(2018)Devlin, Chang, Lee, and Toutanova]{devlin2018bert}
J.~Devlin, M.-W. Chang, K.~Lee, and K.~Toutanova, ``{BERT}: Pre-training of
  deep bidirectional transformers for language understanding,'' \emph{arXiv
  preprint arXiv:1810.04805}, 2018.

\bibitem[Radford et~al.(2019)Radford, Wu, Child, Luan, Amodei, and
  Sutskever]{radford2019language}
A.~Radford, J.~Wu, R.~Child, D.~Luan, D.~Amodei, and I.~Sutskever, ``Language
  models are unsupervised multitask learners,'' 2019.

\bibitem[Microsoft(2020)]{intellisense}
\BIBentryALTinterwordspacing
Microsoft, ``{IntelliSense},''
  \url{https://code.visualstudio.com/docs/editor/intellisense}, 2020. [Online].
  Available: \url{https://code.visualstudio.com/docs/editor/intellisense}
\BIBentrySTDinterwordspacing

\bibitem[JetBrains(2020{\natexlab{a}})]{pycharm}
\BIBentryALTinterwordspacing
JetBrains, ``{PyCharm} code completion,''
  \url{https://www.jetbrains.com/help/pycharm/auto-completing-code.html}, 2020.
  [Online]. Available:
  \url{https://www.jetbrains.com/help/pycharm/auto-completing-code.html}
\BIBentrySTDinterwordspacing

\bibitem[JetBrains(2020{\natexlab{b}})]{intellij}
\BIBentryALTinterwordspacing
------, ``{IntelliJ} code completion,''
  \url{https://www.jetbrains.com/help/idea/auto-completing-code.html}, 2020.
  [Online]. Available:
  \url{https://www.jetbrains.com/help/idea/auto-completing-code.html}
\BIBentrySTDinterwordspacing

\bibitem[Allamanis et~al.(2015)Allamanis, Barr, Bird, and
  Sutton]{allamanis2015suggesting}
M.~Allamanis, E.~T. Barr, C.~Bird, and C.~Sutton, ``Suggesting accurate method
  and class names,'' in \emph{Proceedings of the Joint Meeting of the European
  Software Engineering Conference and the Symposium on the Foundations of
  Software Engineering (ESEC/FSE)}, 2015.

\bibitem[Pratt et~al.(1995)Pratt, Raiffa, Schlaifer,
  et~al.]{pratt1995introduction}
J.~W. Pratt, H.~Raiffa, R.~Schlaifer \emph{et~al.}, \emph{Introduction to
  statistical decision theory}.\hskip 1em plus 0.5em minus 0.4em\relax MIT
  press, 1995.

\bibitem[Mitra et~al.(2018)Mitra, Craswell, et~al.]{mitra2018introduction}
B.~Mitra, N.~Craswell \emph{et~al.}, \emph{An introduction to neural
  information retrieval}.\hskip 1em plus 0.5em minus 0.4em\relax Now
  Foundations and Trends, 2018.

\bibitem[Hellendoorn et~al.(2019)Hellendoorn, Proksch, Gall, and
  Bacchelli]{hellendoorn2019code}
V.~J. Hellendoorn, S.~Proksch, H.~C. Gall, and A.~Bacchelli, ``When code
  completion fails: A case study on real-world completions,'' in \emph{2019
  IEEE/ACM 41st International Conference on Software Engineering (ICSE)}.\hskip
  1em plus 0.5em minus 0.4em\relax IEEE, 2019, pp. 960--970.

\bibitem[Lopes et~al.(2017)Lopes, Maj, Martins, Saini, Yang, Zitny, Sajnani,
  and Vitek]{lopes2017dejavu}
C.~V. Lopes, P.~Maj, P.~Martins, V.~Saini, D.~Yang, J.~Zitny, H.~Sajnani, and
  J.~Vitek, ``D{\'e}j{\`a}vu: a map of code duplicates on github,''
  \emph{Proceedings of the ACM on Programming Languages}, vol.~1, no. OOPSLA,
  p.~84, 2017.

\bibitem[Allamanis(2019)]{allamanis2019adverse}
M.~Allamanis, ``The adverse effects of code duplication in machine learning
  models of code,'' in \emph{Proceedings of the 2019 ACM SIGPLAN International
  Symposium on New Ideas, New Paradigms, and Reflections on Programming and
  Software}, 2019, pp. 143--153.

\bibitem[Miller(1956)]{miller1956magical}
G.~A. Miller, ``The magical number seven, plus or minus two: some limits on our
  capacity for processing information.'' \emph{Psychological Review}, 1956.

\bibitem[Courbariaux et~al.(2014)Courbariaux, Bengio, and
  David]{courbariaux2014training}
M.~Courbariaux, Y.~Bengio, and J.-P. David, ``Training deep neural networks
  with low precision multiplications,'' \emph{arXiv preprint arXiv:1412.7024},
  2014.

\bibitem[Gutmann and Hyv{\"a}rinen(2012)]{gutmann2012noise}
M.~U. Gutmann and A.~Hyv{\"a}rinen, ``Noise-contrastive estimation of
  unnormalized statistical models, with applications to natural image
  statistics,'' \emph{Journal of Machine Learning Research (JMLR)}, 2012.

\bibitem[Attenberg et~al.(2009)Attenberg, Dasgupta, Langford, Smola, and
  Weinberger]{attenberg2009feature}
J.~Attenberg, A.~Dasgupta, J.~Langford, A.~Smola, and K.~Weinberger, ``Feature
  hashing for large scale multitask learning,'' in \emph{Proceedings of the
  International Conference of Machine Learning (ICML)}, 2009.

\bibitem[Moody(1989)]{moody1989fast}
J.~Moody, ``Fast learning in multi-resolution hierarchies,'' in \emph{Advances
  in neural information processing systems}, 1989, pp. 29--39.

\bibitem[Goodfellow et~al.(2016)Goodfellow, Bengio, and
  Courville]{goodfellow2016deep}
I.~Goodfellow, Y.~Bengio, and A.~Courville, \emph{Deep Learning}.\hskip 1em
  plus 0.5em minus 0.4em\relax MIT Press, 2016,
  \href{http://www.deeplearningbook.org}{www.deeplearningbook.org}.

\bibitem[Foundation()]{eclipseCodeRecommenders}
E.~Foundation, ``{Code Recommenders},''
  \href{https://www.eclipse.org/recommenders/}{www.eclipse.org/recommenders},
  visited Mar 2020.

\bibitem[Kuhn and De~Mori(1990)]{kuhn1990cache}
R.~Kuhn and R.~De~Mori, ``A cache-based natural language model for speech
  recognition,'' \emph{Pattern Analysis and Machine Intelligence, IEEE
  Transactions on}, 1990.

\bibitem[Nguyen et~al.(2019)Nguyen, Nguyen, Li, and Wang]{nguyen2019combining}
S.~Nguyen, T.~Nguyen, Y.~Li, and S.~Wang, ``Combining program analysis and
  statistical language model for code statement completion,'' in \emph{2019
  34th IEEE/ACM International Conference on Automated Software Engineering
  (ASE)}.\hskip 1em plus 0.5em minus 0.4em\relax IEEE, 2019, pp. 710--721.

\bibitem[Brockschmidt et~al.(2019)Brockschmidt, Allamanis, Gaunt, and
  Polozov]{brockschmidt2018generative}
M.~Brockschmidt, M.~Allamanis, A.~L. Gaunt, and O.~Polozov, ``Generative code
  modeling with graphs,'' in \emph{Proceedings of the International Conference
  on Learning Representations (ICLR)}, 2019.

\end{thebibliography}

\balance

\end{document}